# Multi-contrast x-ray identification of inhomogeneous materials and their discrimination through deep learning approaches

T. Partridge[1*], S.S. Shankar[2], I. Buchanan[1], P. Modregger[3], A. Astolfo[1,4], D. Bate[4,1], A. Olivo[1]

[1]Department of Medical Physics and Biomedical Engineering, UCL, London, WC1E 6BT, UK
[2]Nylers Ltd, Marshall House, Middleton Road, Morden, Surrey, SM4 6RW, UK
[3]Depatment of Physics, University of Siegen, 57072 Siegen, Germany
[4]Nikon X-Tek Systems Ltd, Tring, Herts, HP23 4JX, UK
[*]corresponding author, tom.partridge@ucl.ac.uk

**Keywords:** X-ray imaging, material discrimination, deep learning

**Abstract**
Recent innovations in x-ray technology (namely phase-based and energy-resolved imaging) offer unprecedented opportunities for material discrimination, however they are often used in isolation or in limited combinations. Here we show that the optimized combination of contrast channels (attenuation at three x-ray energies, ultra-small angle scattering at two, standard deviation of refraction) significantly enhances material identification abilities compared to dual-energy x-ray imaging alone, and that a combination of off-the-shelf machine learning approaches can effectively discriminate e.g., threat materials in complex datasets. The methodology is validated on a range of materials and image dataset that are both an order of magnitude larger than those used in previous studies.
Our results can provide an effective methodology to discriminate, and in some cases identify, different materials in complex imaging scenarios, with prospective applications across the life and physical sciences. While the detection of threat materials is used as a demonstrator here, the methodology could be equally applied to e.g., the distinction between diseased and healthy tissues or degraded vs. pristine materials.

**Significance Statement**
X-ray imaging is an essential tool in many fields, including medical diagnostics, security screening, industrial non-destructive testing and others. New contrast mechanisms are becoming available in x-ray imaging (phase-based imaging, dark field/ultra-small-angle scatter, energy-resolved x-ray imaging) on top of conventional attenuation. These new contrast channels are often used in isolation. In this paper we show that their combined use can be a game-changer in material discrimination and identification (a common goal of many x-ray based inspections), especially if they are used jointly with deep learning approaches. We provide a demonstrator on the discrimination of threat vs benign materials, however the approach is general and can be used across the life and physical sciences.

**Introduction**

The introduction of phase-based methods has transformed x-ray imaging. These methods create contrast based on the unit decrement of the refractive index ($\delta$ if this is written as $n = 1 - \delta + i\beta$ with $i$ the imaginary unit) instead of the imaginary part, $\beta$, which determines the widely used attenuation coefficient by $\mu = 4\pi\beta/\lambda$ (with $\lambda$ the x-ray wavelength). Following pioneering experiments in the 1960s (1) and 1980s (2), the field exploded in the 1990s with the advent of 3rd generation synchrotron facilities (3-4) and with the first pioneering experiments based on more conventional micro-focal sources (5-6).

Effort was soon focused on approaches to perform quantitative phase imaging by separating attenuation and phase contributions (phase retrieval), with methods based both on crystals (7) and free-space propagation (8) being developed. Crystal-based methods (as well as methods developed later based on x-ray masks (9), gratings (10) or other beam modulators (e.g., 11-13)) are actually sensitive to x-ray refraction i.e., to the first derivative of the phase shifts, which is linked to the refraction angle $\alpha$ by $\alpha = \frac{\lambda}{2\pi}\nabla_{x,y}\phi(x, y, \lambda)$ with $\nabla_{x,y}$ the two-directional gradient operator transverse to the x-ray propagation direction $z$, and $\Phi$ the phase shift introduced by the object. For this reason, these approaches are referred to as "differential phase" methods; if needed, the phase shift $\Phi$ can be obtained by integration.

Experimentation with crystal-based methods soon revealed that a third contrast channel could be retrieved, related to multiple refraction events caused by structures smaller than the spatial resolution of the imaging system. Early papers called this contrast channel refractive scattering (14), extinction (15) or borrowed the term ultra-small angle x-ray scattering (USAXS) from the x-ray scattering community (16), although it must be borne in mind that in this case angles are in the order of microradians rather than degrees or tenths of a degree. Later, the community converged on the name dark-field (17), although USAXS is still widely used. The dark-field channel turned out to be accessible through practically all differential phase methods (17-19), including those adapted for use with low-brilliance x-ray sources in the 2000s (20-21), with options to access it also through propagation-based approaches emerging only more recently (22). In addition to enabling some of the earliest clinical trials with phase-based x-ray methods (23), access to dark-field alongside differential phase with low-brilliance sources makes "multi-contrast" (phase, attenuation, dark-field) x-ray imaging readily available in standard labs, opening the way to commercial translation.

Another transformative development in x-ray imaging was the introduction of energy-resolved detectors (24-25). These allow splitting the incoming x-ray spectrum in two (or more) parts by means of opportunely calibrated energy thresholds; their use in phase-based multi-contrast imaging therefore makes the above three channels simultaneously available in high and low energy versions, making six contrast channels available in principle. Furthermore, with mask-based methods like the one used in this work, the x-rays transmitted through the highly absorbing mask septa can also be collected separately to create a third attenuation image at a much higher x-ray energy (26). Clearly, these seven contrasts are not all independent from each other. The independence of low and high energy x-ray attenuation images can be assumed when these are dominated by the photoelectric and the Compton effects, respectively (27). More recently, a different energy dependence has also been observed in gratings-based dark field imaging for features with sizes above and below a certain characteristic length of the imaging system (28). Conversely, phase scales in a predictable way with x-ray energy, and its correlation with the Compton signal has been widely discussed (e.g., 29). In the framework of attenuation-based imaging, it has also been repeatedly observed that the problem over-determination allowed by the availability of attenuation images at more than two energies leads to an increased precision in the results (e.g., 30), and that similar principles can be extended to phase-based imaging (26, 31), which this paper expands on to explore options for material identification based on the simultaneous availability of multiple image contrasts.

As a demonstrator, we apply the method to the area of security inspections, namely to the identification of threat materials. The field of X-ray based security scans is well developed, with exhaustive summaries provided in various review papers (e.g., 32). While long-established dual-energy scanners are still widely used (33), technologies such as Computed Tomography (CT) are being deployed at airports worldwide, while others such as x-ray diffraction (e.g., 34) are experiencing early market entry. However, dark-field imaging probes a different property of matter, namely microstructure instead of atomic number/electron density and molecular structure (probed by CT and diffraction, respectively), therefore providing information that is complementary to existing methods. Furthermore, dark-field is provided simultaneously with attenuation, and therefore on top of the currently used dual-energy images if an energy-resolved detector or appropriate filtration is used.

The combination of dark-field with deep learning approaches is also explored in this work. The popularity of deep learning methods has been booming in recent years (35), and their use in security applications is also rapidly expanding (36). The prospect of automated detection is particularly attractive in security inspections, and indeed 3D imagery like that provided by CT is particularly well suited for analysis via machine learning methods (37). Here we are interested in exploring the compatibility between multi-contrast x-ray imaging and deep learning, in particular the added value brought by the dark-field channel.

In the first part of the paper, we test the discrimination and identification potential of multi-contrast x-ray imaging used as a standalone analytic method with no added automatic detection algorithms. We apply this to a large dataset encompassing 19 threat and 56 non-threat materials of varying thickness, showing how the inclusion of additional contrast channels above conventional attenuation significantly aids material discrimination and often allows material identification. We propose a method to convert multi-contrast images into simpler, "material-specific" images and analyze the overall results obtained using this approach by means of truth tables. We extend the methodology to the case where different materials overlap, showing the approach still works.

In the second part of the paper, we create complex scenarios where the 75 materials are hidden inside bags and covered by a range of cluttering materials. Overall, we produced 3891 scans of randomly selected materials, with the thickness of each material varying between 12 and 30 mm in the various scans, combined with various cluttering objects. 2732 (70%) of these scans were used as a training set (for transfer learning), and the remaining 1168 (30%) for testing. We progressively optimized our deep learning approach, and obtained the best results with a hierarchical approach where the cluttering objects are segregated first then materials are discriminated; this was trained with a cross-entropy loss function and used layer-specific learning rate adaptation (38). Despite the complexity of the dataset, use of this architecture resulted in a single false negative out of the 313 explosive-bearing cases contained in the set of 1168 images used for testing.

We note that, although here we used the discrimination of threat materials as a demonstrator, our interest lays in the general ability of dark field to separate, and ideally identify materials using additional information regarding their microstructure.

**Results and discussions**
An initial simplified example of the benefit of using additional contrast channels alongside conventional ones is shown in Fig. 1. The horizontal axis in this plot is the effective atomic number ($Z_e$) corrected using published linear attenuations (39), and represents a condensed version of dual energy attenuation-based imaging. The additional contrast channels accessed through phase contrast measurement with masks are then added on the vertical axis. These are dark field at low (a) and high (b) energy, the "offset" i.e., the image at a much higher average x-ray energy obtained by exploiting x-rays that have traversed the highly absorbing mask septa (c), and the standard deviation of the refraction signal (d), used as a proxy for dark field when features with a size comparable to or slightly larger than the system's spatial resolution are present (40). For a straightforward comparison with $Z_e$, all contrast channels bar the standard deviation of refraction have been made thickness-independent through division by the attenuation signal, following the procedure described in (26). Explosive materials are highlighted with a square for ease of visualization.

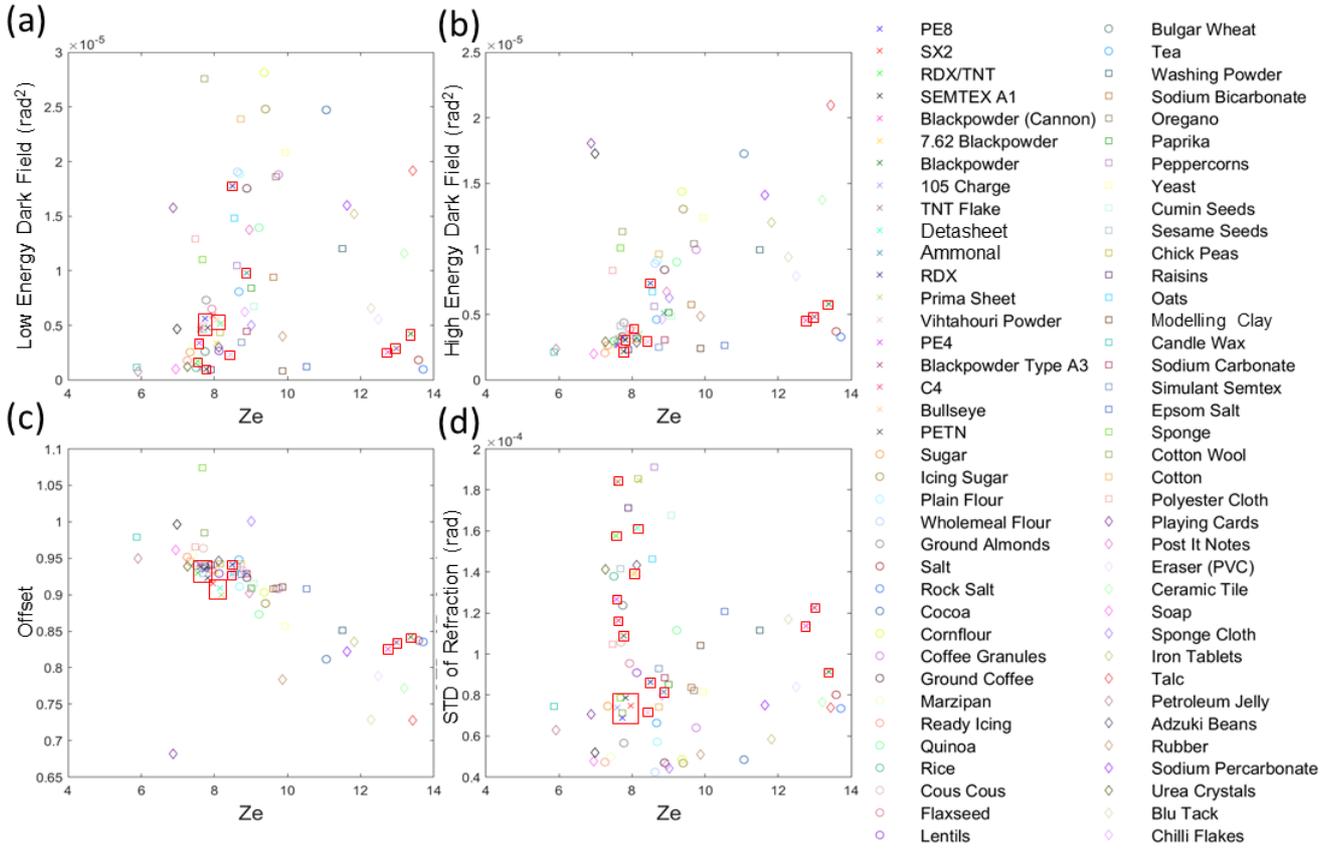

*Fig. 1 Scatterplots combining the effective atomic number ($Z_e$, horizontal axis in all graphs) with dark field at low and high energy, offset and standard deviation of refraction, reported on the vertical axis of panels (a), (b), (c) and (d), respectively. Explosive materials are highlighted by squares.*

As can be seen, adding contrast channels separates out the points in the 2D graphs, allowing for much easier identification – also and especially of materials that would overlap based on their $Z_e$ value alone. Unsurprisingly, the offset signal is the least effective in providing additional separation, due to its lack of complementarity over the two energies used to extract $Z_e$. Dark field at low and high energy provide similar results, with the former spreading out the points a bit more due to the dark field signal being stronger at lower energies (28). The best separation seems to be provided by the standard deviation of refraction, which would indicate a prevalence of grain sizes comparable to or larger than the spatial resolution of the imaging system (approximately 25 μm).

From an operational perspective, the above information needs to be combined in a simplified way that could be presented to e.g., an operator, without trading off on signal complexity. We propose a simple quantitative approach based on the relative distance between material pairs in all contrast channels, exemplified in Fig. 2. A radius corresponding to the measurement error (e.g., one standard deviation over a region of interest (ROI)) is assigned to all reference points (e.g., blue arrows in Fig 2 for PE8). When an unknown material is scanned, contrast measurements and corresponding standard deviations are extracted from the corresponding ROI, and plotted on top of the reference materials (red arrows in Fig 2). While for obvious reasons only two contrast channels are represented in figure 2, this procedure actually defines a "hyper-cube" (or hyper-sphere if standard deviations are summed in quadrature), with dimensions corresponding to the number of used contrast channels. This immediately provides the probability of an unknown sample being made of a certain material as the overlap between hyper-cubes (hyper-spheres).

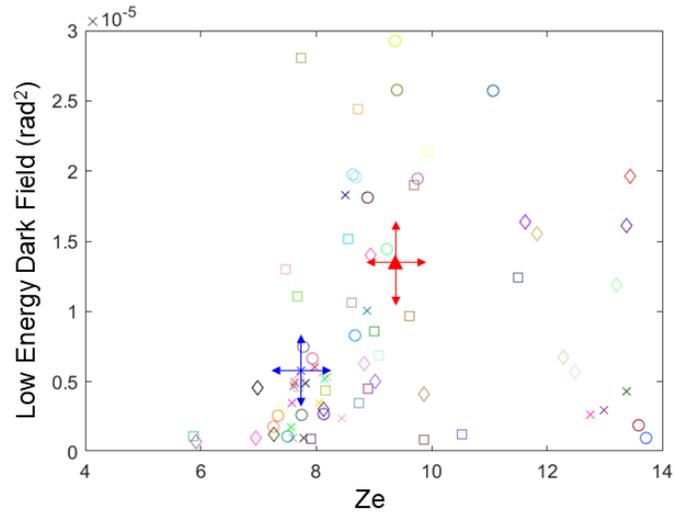

*Fig. 2 material discrimination based on relative distances in "image contrast" space, with the overlap between areas defined by the uncertainty on the measured contrasts determining the probability of an unknown (red) material corresponding a known one (blue). While only two contrasts can be represented in a 2D figure, in truth the overlap between "hyper-volumes" is calculated by simultaneously considering the uncertainties on all contrast channels.*

Those probabilities can be used to produce user-friendly, "material-specific" images, an example of which is provided in figure 3; the sides(radii) of the hyper-cubes(spheres) can be used to trade off sensitivity vs specificity.

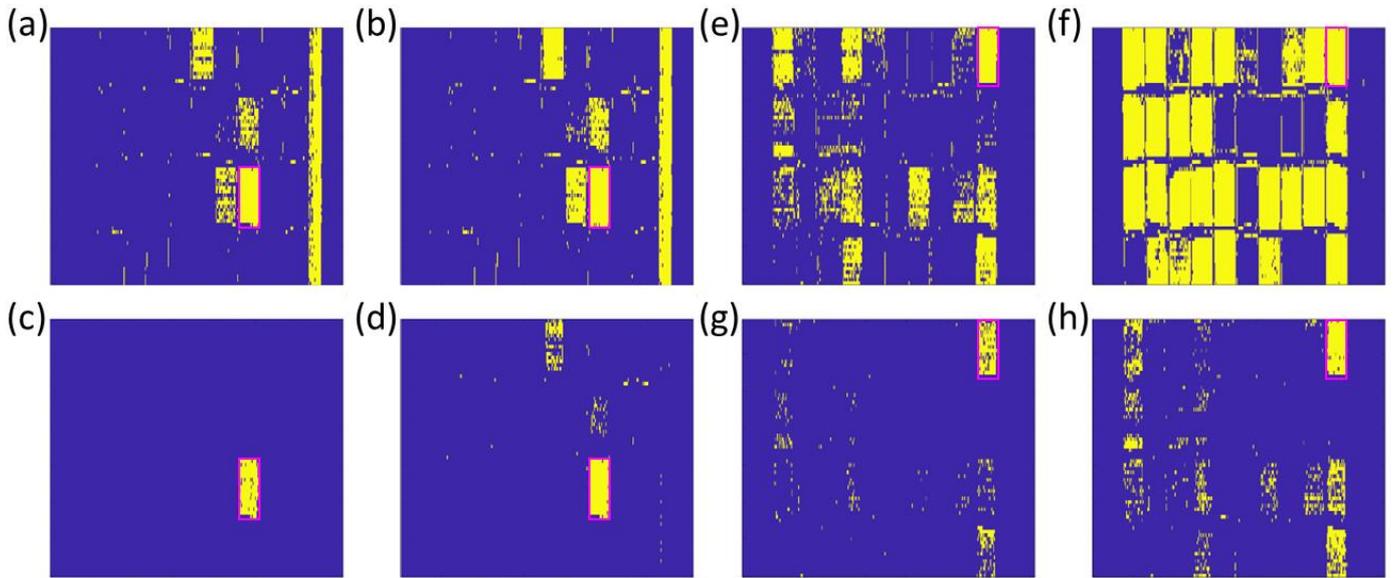

*Fig 3 Material-specific images for bicarbonate of soda (a-d) and semtex (e-h), the area corresponding to which is highlighted by a pink rectangle in all images. All images are binary, flagging the pixels as belonging to the material in question (yellow) or not (blue). Images (a-b and e-f) are based only on $Z_e$, with short (a, e) and long (b, f) cutoff radii corresponding to high specificity and high sensitivity conditions, respectively. Images (c-d and g-h) include the additional contrast channels introduced in Fig. 1. The plurality of contrasts provides different mechanisms to trade-off specificity and sensitivity; for example, in images c and g (high specificity), $Z_e$ had to be triggered alongside 3 of the other 4 contrast channels for a pixel to be flagged as belonging to a certain material. Conversely, "high sensitivity" images in d and h simply require that 3 contrasts out of 5 are triggered, regardless of $Z_e$. The improvement brought by the proposed multi-contrast method can be appreciated by comparing each image of the bottom row to its "$Z_e$ only" correspondent immediately above.*

The examples in Fig. 3 refer to bicarbonate of soda (a-d) and semtex (e-h), respectively. (a-b) and (e-f) use $Z_e$ only, with a smaller (a, e) and larger (b, f) thresholds resulting in high specificity and sensitivity, respectively. Images (c-d) and (g-h) introduce dark-field at two energies, offset and standard deviation of refraction. Multiple contrast channels lead to multiple ways to trade-off sensitivity and specificity: for example, highly specific images (c) and (g) require that $Z_e$ is triggered alongside three of the other four contrast channels, while (d) and (h) simply require that three channels are flagged, one of which may or may not be $Z_e$. As can be seen the multi-contrast approach works extremely well for bicarbonate of soda, where it completely clears up the image; relaxing the specificity makes almost all pixels in the bicarbonate ROI turn yellow, at the expense of some spurious positive pixels in other parts of the image. Things work slightly less well for semtex, however the combination of contrasts significantly outperforms the corresponding $Z_e$ only cases. In future work, outcomes could be further improved by a) giving different weights to the various contrast channels, which could be refined with successive and extensive calibrations and b) performing additional probability calculations based on the number of pixels flagged in a ROI, e.g., by assuming "continuity" for the examined material and defining likely boundaries. Supplementary figure 1 provides additional examples for different materials, and Supplementary figures 2 and 3 show that the method is still applicable when different materials overlap.

Application of the above principle to a range of different materials allows the creation of confusion matrices, an example of which is provided in Fig. 4.

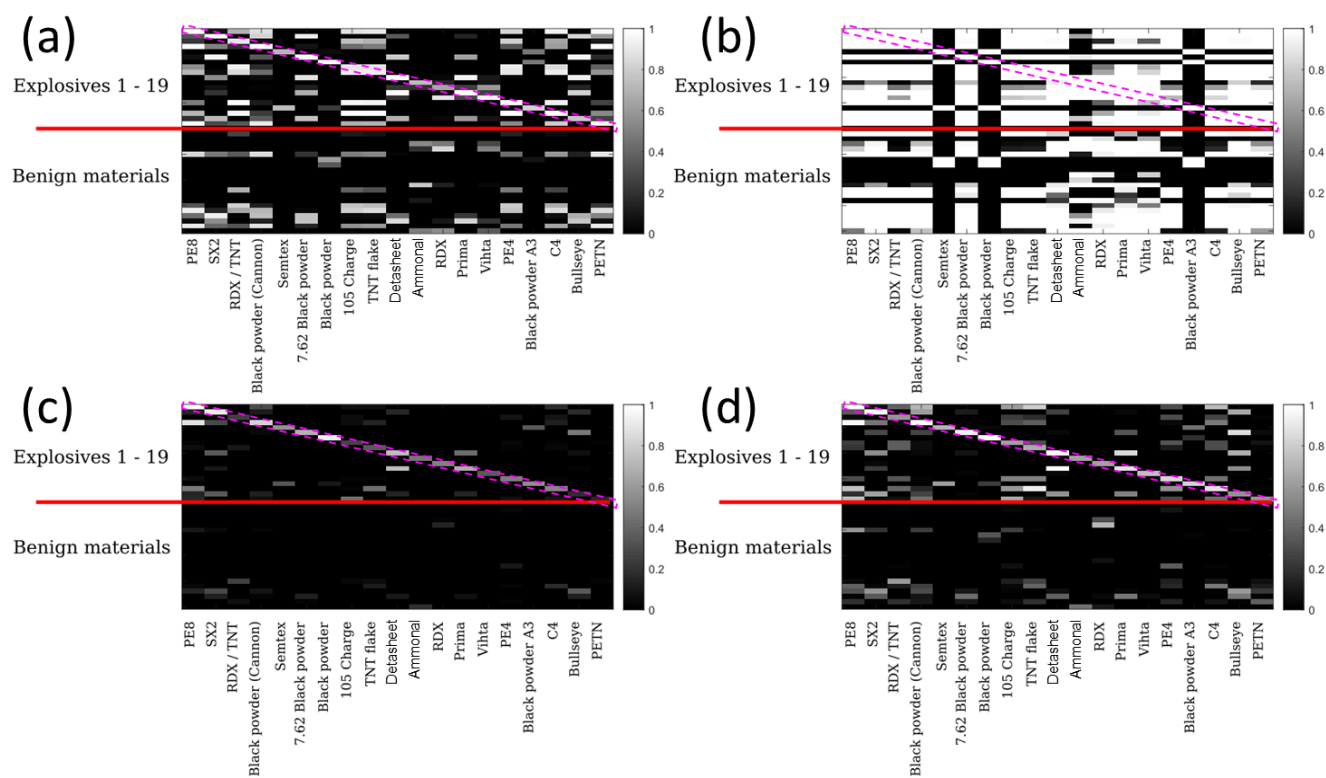

*Fig. 4. Confusion matrices based on $Z_e$ at high specificity (a) and high sensitivity (b), and on the multi-contrast approach at high specificity (c) and high sensitivity (d). Matrices focus only on the explosive materials, which are listed on the horizontal axis. The same materials in the same order occupy the top 19 rows of each matrix, while the bottom 19 represent a subset of the benign materials. The grey scale represents the probability that a certain material belongs to a certain class, hence the ideal matrix is a white diagonal occupying the top half of the graph, with all other entries being black.*

Confusion matrices like those of Fig. 4 enable presenting results like those of Fig. 3 for a plurality of materials in a single image. The grey scale represents the probability that a material belongs to a certain class, with white and black representing certainty of belonging/not belonging to that class, respectively. The same materials are listed along the horizontal and vertical axis, so that a white diagonal in a black background represents perfect detection with 100% certainty (here we have represented twice as many materials on the vertical than on the horizontal axis, so the diagonal would only occupy half the graph). As can be seen, the $Z_e$ cases in (a) and (b) are severely plagued by false positives, also in the high specificity case. The multi-contrast analysis significantly clears up the confusion matrix; it must be noted, however, that the diagonal points in (c) are dimmer than in (a) and (b). This is ameliorated by relaxing the specificity in favor of sensitivity (d), which results in the appearance of a larger number of false positives. Overall, (d) has a comparable diagonal to (a), but fewer (and dimmer) false positives.

Finally, we present the results of a test carried out using deep learning. For simplicity, we focused on the performance of attenuation at high and low energy alone vs. their combination with dark-field at high and low energy. A random combination of materials with random thickness ranging from 12 to 30 mm was placed inside bags with additional cluttering items; ratio images were used to eliminate the dependence from material thickness as described in (26). 3891 combinations were produced and scanned with the imaging system described in the methods section. Supplementary table 1 lists all threat and non-threat materials as well as all cluttering items, supplementary figure 4 provides a scheme summarizing the characteristics of the dataset used in the deep learning experiments, supplementary figure 5 shows a few example images from the scanning campaign.

The 3891 images were split into 2732 (70%) for training and 1168 (30%) for testing. The convolutional neural network (CNN) Inception V3 (41) was used as our base architecture, pre-trained on ImageNet (42); supplementary figure 6 shows a schematic of our hierarchical architecture (that yielded the best performance), and more details on it are provided in the methods section. We progressively refined our approach, by 1) moving from a single to a hierarchical architecture in which the cluttering object classes are segregated first then the materials differentiated, 2) swapping the softmax for a cross-entropy loss function, and 3) implementing layer-specific learning rate adaptation (38): every step increased the obtained precision and recall values. To calculate these, we focused on the 313 images out of the 1168 used for testing that contained an explosive. We counted the number of true positives (TP), false positives (FP) and false negatives (FN), and calculated precision as TP/(TP+FP) and recall as TP/(TP+FN); note that TP + FN = 313 while typically TP + FP > 313 as FPs are images outside the 313 explosive-containing ones erroneously assumed to contain an explosive. Precision and recall provide estimates of specificity and sensitivity, respectively; results are reported in table 1 for the various architecture refinements. The best performing architecture (hierarchical segregation + cross-entropy loss function + layer-specific learning rate adaptation) was also applied to a dataset from which the dark-field images had been eliminated, resulting in a significantly reduced performance.

| Architecture | TP | FP | FN | Precision | Recall |
|---|---|---|---|---|---|
| Single softmax | 297 | 65 | 16 | 82.04% | 94.89% |
| Hierarch. softmax | 303 | 57 | 10 | 84.17% | 96.81% |
| Hierarch. x-entropy | 307 | 52 | 6 | 85.52% | 98.08% |
| Hierarch. x-entropy +layer-specific learn. adapt. | 312 | 32 | 1 | 90.70% | 99.68% |
| Hierarch. x-entropy +layer-specific learn. adapt. **no dark field** | 281 | 56 | 32 | 83.38% | 89.78% |

*Table 1 Results of machine learning analysis performed with various architectures. The first four rows refer to datasets combining attenuation at two energies and dark field at two energies, while the two dark-field images were removed to obtain the results of row 5. TP = true positives, FP = false positives, FN = false negatives, Precision = TP/(TP+FP), Recall = TP/(TP+FN).*

The table clearly shows the improvement in both precision and recall values as the deep learning architecture is progressively refined, reaching a near-perfect 99.68% degree of recall for the hierarchical architecture with cross-entropy loss function and layer-specific learning adaptation; note this corresponds to a single false negative out of the 313 explosive-bearing images included in the 1168 used for testing. Importantly, removing the dark-field imaging channel from the dataset leads to a significant degradation in the overall performance, with a reduction of 7 and 10 percentage points in the precision and recall values, respectively, thereby demonstrating the importance of the dark-field channel in the material discrimination potential of deep neural networks.

**Conclusions**
We have shown that the inclusion of additional x-ray contrast channels, namely dark-field at two average energies, attenuation at a third energy and standard deviation of refraction can significantly enhance the material discrimination and identification potential of dual-energy x-ray imaging. We developed a mechanism to combine all the different contrast channels into a single "material-specific" image, the pixels of which represent whether they belong to that material based on overlapping hyper-spheres, the radii of which are a chosen uncertainty. We propose a mechanism to trade-off sensitivity and specificity in this approach and compare the results to those obtained while using only attenuation at two x-ray energies, demonstrating substantial advantages. We also provide evidence that the method can be extrapolated to overlapping materials, so long as areas where we can "re-normalize" the signal against the overlapping object/background are available.

This laid the groundwork for the application of deep learning to the multi-contrast images, which was previously shown to bear promise in a proof-of-concept study involving a small dataset and a limited number of materials (26). In this wider study, a custom-developed architecture was applied to almost 4000 images containing 57 benign and 19 explosive materials with varying thickness, randomly mixed in bags also containing a multitude of cluttering objects. The network architecture was progressively refined by creating a hierarchical structure in which the cluttering object classes are segregated first then materials are discriminated, as well as by introducing a cross-entropy loss function and layer-specific learning rate adaptation; this ultimately led to a near-perfect recall rate of 99.68%, corresponding to a single explosive being missed out of the 313 contained in the 1168 images used for testing. By applying the same deep neural network to the same image set from which the dark-field images were excluded, the recall rate dropped by 10 percentage points, corresponding to 32 explosives being missed. We find this highlights the fundamental importance of the match between dark field images and detection abilities of the deep neural network in discriminating certain material classes. Although security inspections have been chosen as an example to demonstrate the potential of the approach, the method is purely based on the additional discrimination capabilities provided by the textural nature of dark-field images, related to the microscopic structure of a given material; as such, it can be widely applied to microscopically inhomogeneous materials across the life and physical sciences.

**Methods**
*X-Ray imaging system and method:* this work used the edge-illumination (EI) x-ray phase-contrast imaging method (9) in its laboratory implementation (21) to produce the dark-field and refraction images on top of the conventional attenuation ones. Combination with an x-ray detector with two energy thresholds (XCounter XC-FLITE FX2, Direct Conversion, Danderyd, Sweden) allowed acquiring all the above images at two different average spectral energies. The x-ray source (MXR-160HP/11, Comet, Wünnewil-Flamatt, Switzerland) was set at 120 kVp, and a higher detector threshold roughly splitting the (W) spectrum in half (i.e., with approximately equal number of counts in the high and low energy bins) was selected. X-ray energies below approximately 18 keV were cut off by the lower detector threshold, so as to completely eliminate noise (i.e., detector pixels registered no counts with the beam off).

Two masks were used to realize the EI configuration. The first ("pre-sample") mask, placed immediately before the sample, had overall dimensions of 150 (h) x 9.6 (w) mm$^2$, with regularly spaced (75 μm period), 21.4 μm apertures parallel to the long side of the mask. The second ("detector") mask, placed immediately before the detector, had dimensions of 200 (h) x 12.8 (w) mm$^2$, 98 μm period, 28 μm apertures. The source-to-detector distance was 2.1 m; the detector mask was placed at 42 mm from the detector, so that its projected period would match the 100 μm detector pixel pitch. The pre-sample mask was slightly displaced from its "ideal" position at 52.5 cm from the detector, where it also would also have matched the detector pitch; in this way, the beamlets it creates hit slightly different positions on the detector mask (see Fig. 5), allowing for the retrieval of attenuation, refraction, and dark field with a single sample scan, similarly to the method described in (23). Basically, different detector pixels receive different amounts of illumination (Fig 5(a)), which corresponds to sampling the illumination curve (IC) at different points, thus allowing its fit with an appropriate mathematical function. The IC is obtained by scanning the pre-sample mask while the detector mask is kept stationary, with apertures at the center of the corresponding pixels. This provides a bell-shaped curve with the maximum where pre-sample and detector mask apertures are aligned, and minima where the pre-sample mask apertures are aligned with the septa at either side of the corresponding detector masks aperture. The introduction of a sample causes the IC to dampen, shift laterally, and broaden, corresponding to attenuation, refraction, and dark-field effects, respectively. Since the IC is normally well approximated by a Gaussian function (18, 26, 43), fitting it before and after the introduction of a sample enables extracting the latter's attenuation, refraction, and dark-field characteristics on a pixel-by-pixel basis. Specifically, if:

$$IC_0 = \frac{a_0}{\sqrt{2\pi a_2}} exp\left[\frac{(x-a_1)^2}{2a_2}\right] + a_3 \qquad (1)$$

is the IC without the object, in which a term ($a_3$) accounting for a degree of x-ray transmission through the mask septa has also been included, and:

$$IC_{obj} = \frac{a_0 t}{\sqrt{2\pi(a_2+\sigma)}} exp\left[\frac{(x-a_1-\Delta)^2}{2(a_2+\sigma)}\right] + a_3 o \qquad (2)$$

is the IC with the object, then $t$, $\Delta/d$, $\sigma/d^2$ and $o$ represent the transmission, refraction, dark-field and offset images, respectively, with $d$ the sample to detector mask distance used to transform the lateral shift of the IC into the refraction angle $\vartheta$, in the approximation $\tan(\vartheta) \simeq \vartheta$.

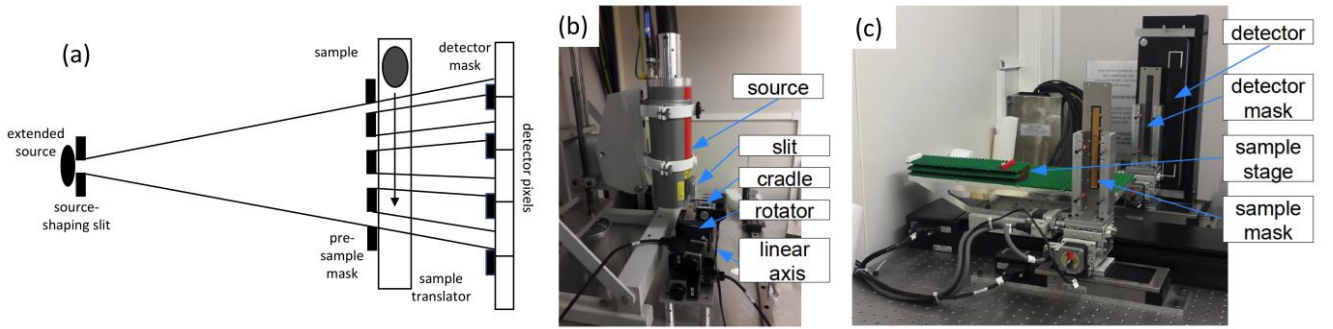

*Figure 5 schematic of the experimental setup (a), with photos of the source (b) and detector/masks (c) parts. A mismatch of the pre-sample mask position along the x-ray propagation axis allows for different detector apertures being hit at different points by the beamlets created by the pre-sample mask.*

Both masks are mounted on a pair of translators and a goniometer for alignment. An additional, larger translator is used to scan the samples through the beam. The relatively large (0.4 mm) focal spot of the source is reduced to approximately 80 μm in the horizontal direction by means of a Huber (Huber Diffraktionstechnik GmbH & Co. KG, Rimsting, Germany) slit, placed as close as possible to the source output window (see Fig. 5(b)). A detailed description of the system can be found in (43), save for the use of a slightly different pre-sample mask. The entire system is placed inside a cabinet, which enables its easy transportation to different locations (e.g., for the explosive scans, see below).

*Samples and sample preparation:*
A complete list of the materials considered in this application is provided in supplementary table 1, which also lists the cluttering objects used in the deep learning part of the study alongside the 57 benign and 19 explosive materials. The benign materials were scanned on the Nikon X-Tek Systems premises in Tring (Hertfordshire, UK) using the pre-commercial prototype described above. The prototype was then transported to the Cranfield Ordnance Test and Evaluation Centre (COTEC) near Devizes (Wiltshire, UK), which provided access to the explosive materials. These were first scanned individually as done for the benign materials in Tring, then placed into bags and obscured by cluttering objects. A total of 3891 different combinations were produced and scanned, a description of which is provided in Supplementary Figure 4, which also shows what combination of cluttering objects were used in combination with which material. At least 40 (typically more) cluttered combinations were realized for each of the threat materials, which means the overall number of scans containing an explosive was approximately 22% of the total.

*Deep learning architectures:*
The CNN Inception V3 (41) was used as our base architecture. This was pre-trained on ImageNet (42), with a few additional layers trained on our (training) dataset. For training, we used stochastic gradient descent with momentum as the optimization procedure, with L2 as the regularizer (43). We trained for 20 epochs with a batch size of 64, using a dropout probability of 0.5 for the convolutional layers. A 10 times lower learning rate was used for the pre-trained layers in comparison to the additional ones, using batch normalization (45) after the convolutional layers of Inception V3. Initially we used the softmax loss function, then introduced a cross-entropy loss. Assume we have $N$ training samples, with $M$ classes. Let the training set be represented by $\{(x_1, y_1),…, (x_N, y_N)\}$ where $x_i, i \in \{1, ..., N\}$ represent $N$ training images, and $y_i$ denotes the corresponding ground-truth labels (since we have $M$ classes, $y_i \in \{1, ..., M\}$). Training with a softmax loss is accomplished by minimizing the negative log-likelihood:

$$\mathcal{L}_s = -\frac{1}{N} \sum_{r=1}^{N} \log(\hat{p}_{r,y_r})$$

where the probability $\hat{p}_{r,y_r}, r \in \{1, ..., N\}$ is obtained by applying the softmax function to the penultimate layer of the classifier. Letting $l_{r,m}$ denote the m$^{th}$ output for $x_r$, we have:

$$p'_{r,m} = \frac{e^{l_{r,m}}}{\sum_{m'} e^{l_{r,m'}}}, \quad m, m' \in \{1, ...., M\}$$

Conversely, with a sigmoid cross-entropy loss, the network is trained by minimizing the following loss function:

$$\mathcal{L}_e = -\frac{1}{NM} \sum_{r=1}^{N} [\boldsymbol{p}_r \cdot \log(\hat{\boldsymbol{p}}_r)] + [(1 - \boldsymbol{p}_r) \cdot \log(1 - \hat{\boldsymbol{p}}_r)]$$

with $\boldsymbol{p}_r$ the $M$-dimensional ground truth probability vector, and $\hat{\boldsymbol{p}}_r$ the $M$-dimensional predicted probability vector obtained by applying the sigmoid function to the outputs of the penultimate layer of the classifier.

While progressively refining the approach, the first improvement consisted of the introduction of a hierarchical architecture, to segregate cluttering object classes first, then apply material discrimination. Assuming we have $K$ object classes and $R$ materials, an inception V3 net is first trained to detect one of the $K$ objects $O_1….O_K$. $K$ more inception V3 nets are then separately trained, where the $l^{th}$ net ($1 \leq l \leq K$) separates the benign and threat materials obscured by $O_l$. If a material image has been obscured by two objects $O_{k_1}$ and $O_{k_2}$, the image is put into both $O_{k_1}$ and $O_{k_2}$ object classes during training with mild data augmentation (see Supplementary Figure 4). During testing, the trained model chooses the $k_1$-th material-discrimination net if the probability of the presence of object $O_{k_1}$ is predicted to be greater than that of the presence of object $O_{k_2}$, and vice versa. This functionality naturally extends if a material is obscured by three or more objects.

The second improvement consisted of the introduction of the sigmoid cross-entropy loss function outlined above; note that the change of loss function only applied to the $K$ nets trained for material discrimination, while the first net is still trained with the softmax loss. Finally, we introduced layer-specific learning rate adaptation, by assigning a weight to a given layer proportional to the average decorrelation-based segregation between classes (38); again, this was only applied to the $K$ material-discrimination nets.

Training and inference procedures were carried out on an NVIDIA GeForce GTX 1080 Ti graphic card, with on-chip memory of ~11 GB, which led to an inference time per image of the order of 10 ms.

**References**

1. U. Bonse, M. Hart, An x-ray interferometer. *Appl. Phys. Lett.* **6**, 155-6 (1965).
2. E. Förster, K. Goetz, P. Zaumseil, Double crystal diffractometry for the characterisation of targets for laser fusion experiments. *Kristall Tech.* **15**, 937-45 (1980).
3. A. Snigirev, I. Snigireva, V. Kohn, S. Kuznetsov, I. Schelokov, On the possibilities of x-ray phase-contrast microimaging by coherent high-energy synchrotron radiation. *Rev. Sci. Instrum.* **66**, 5486-92 (1995).
4. A. Momose, T. Takeda, Y. Itai, Y, K. Hirano, Phase-contrast X-ray computed tomography for observing biological soft tissues. *Nat. Med.* **2**, 473-5 (1996).
5. T.J. Davis, D. Gao, T.E. Gureyev, A.W. Stevenson, S.W. Wilkins, Phase-contrast imaging of weakly absorbing materials using hard x-rays. *Nature* **373**, 595-8 (1995).
6. S.W. Wilkins, T.E. Gureyev, S. Gao, A. Pogany, A.W. Stevenson, Phase-contrast imaging using polychromatic hard x-rays. *Nature* **384**, 335-8 (1996).
7. D. Chapman, W. Thomlinson, R.E. Johnston, D. Washburn, E. Pisano, N. Gmür, Z. Zhong, R. Menk, F. Arfelli, D. Sayers, Diffraction enhanced x-ray imaging. *Phys. Med. Biol.* **42**, 2015-25 (1997).
8. P. Cloetens, W. Ludwig, J. Baruchel, D. Van Dyck, J. Van Landuyt, J.P. Guigay, M. Schlenker, Holotomography: quantitative phase tomography with micrometer resolution using hard synchrotron radiation x-rays. *Appl. Phys. Lett.* **75**, 2912-4 (1999).
9. A. Olivo, F. Arfelli, G. Cantatore, R. Longo, R.H. Menk, S. Pani, M. Prest, P. Poropat, L. Rigon, G. Tromba, E. Vallazza, E. Castelli, An innovative digital imaging set-up allowing a low-dose approach to phase contrast applications in the medical field. *Med. Phys.* **28**, 1610-9 (2001).
10. C. David, B. Nohammer, H.H. Solak, Differential x-ray phase contrast imaging using a shearing interferometer. *Appl. Phys. Lett.* **81**, 3287-9 (2002).
11. K.S. Morgan, D.M. Paganin, K.K.W. Siu, Quantitative single-exposure x-ray phase contrast imaging using a single attenuation grid. *Opt. Exp.* **19**, 19781-89 (2011).
12. K.S. Morgan, D.M. Paganin, K.K.W. Siu, X-ray phase imaging with a paper analyser. *Appl. Phys. Lett.* **100**, 124102 (2012).
13. S. Berujon, H. Wang, K. Sawhney, X-ray multimodal imaging using a random-phase object. *Phys. Rev. A* **86**, 063813 (2012).
14. L. Rigon, H.J. Besch, F. Arfelli, R.H. Menk, G. Heitner, H. Plotow-Besch, A new DEI algorithm capable of investigating sub-pixel structures. *J. Phys. D: Appl. Phys.* **36**, A107-12 (2003).
15. O. Oltulu, Z. Zhong, M. Hasnah, M.N. Wernick, D. Chapman, Extraction of extinction, refraction and absorption properties in diffraction enhanced imaging. *J. Phys. D: Appl. Phys.* **36**, 2152-6 (2003).
16. E. Pagot, P. Cloetens, S. Fiedler, A. Bravin, P. Coan, J. Baruchel, J. Härtwig, W. Thomlinson, A method to extract quantitative information in analyzer-based x-ray phase contrast imaging. *Appl. Phys. Lett.* **82**, 3421-3 (2003).
17. F. Pfeiffer, M. Bech, O. Bunk, P. Kraft, E.F. Eikenberry, C. Brönnimann, C. Grünzweig, C. David, Hard-x-ray dark-field imaging using a grating interferometer. *Nat. Mat.* **7**, 134-7 (2008).

18. M. Endrizzi, P.C. Diemoz, T.P. Millard, J.L. Jones, R.D. Speller, I.K. Robinson, A. Olivo, Hard x-ray dark-field imaging wit incoherent sample illumination. *Appl. Phys. Lett.* **104**, 024106 (2014).
19. I. Zanette, T. Zhou, A. Burvall, U. Lundström, D.H. Larsson, M. Zdora, P. Thibault, F. Pfeiffer, H.M. Hertz, Speckle-based x-ray phase-contrast and dark-field imaging with a laboratory source. *Phys. Rev. Lett.* **112**, 253903 (2014).
20. F. Pfeiffer, T. Weitkamp, O. Bunk, C. David, Phase retrieval and differential phase-contrast imaging with low-brilliance x-ray sources. *Nat. Phys.* **2**, 258-61 (2006).
21. A. Olivo, R. Speller, A coded-aperture technique allowing x-ray phase contrast imaging with conventional sources. *Appl. Phys. Lett.* **91**, 074106 (2007).
22. T.E. Gureyev, D.M. Paganin, B. Arhatari, S.T. Taba, S. Lewis, P.C. Brennan, H.M. Quiney, Dark-field signal extraction in propagation-based phase-contrast imaging. *Phys. Med. Biol.* **65**, 215029 (2020).
23. K. Willer, A.A. Fingerle, W. Noichl, F. De Marco, M. Frank, T. Urban, R. Schick, A. Gustschin, B. Gleich, J. Herzen, T. Koehler, A. Yaroshenko, T. Pralow, G.S. Zimmermann, B. Renger, A.P. Sauter, D. Pfeiffer, M.R. Makowski, E.J. Rummeny, P.A. Grenier, F. Pfeiffer, X-ray dark-field chest imaging for detection and quantification of emphysema in patients with chronic obstructive pulmonary disease: a diagnostic accuracy study. *Lancet Digital Health* **3**, e733-44 (2021).
24. E. Roessl, R. Proksa R, K-edge imaging in x-ray computed tomography using multi-bin photon counting detectors. *Phys. Med. Biol.* **52**, 4679-96 (2007).
25. D. Pacella, Energy-resolved x-ray detectors: the future of diagnostic imaging. *Rep. Med. Imaging* **8**, 1-13 (2015).
26. T. Partridge, A. Astolfo, S.S. Shankar, F.A. Vittoria, M. Endrizzi, S. Arridge, T. Riley-Smith, I.G. Haig, D. Bate, A. Olivo, Enhanced detection of threat materials by dark-field x-ray imaging combined with deep neural networks. *Nat. Commun.* **13**, 4651 (2022).
27. R.E. Alvarez, A. Macovski, Energy-selective reconstructions in x-ray computerized tomography. *Phys. Med. Biol.* **21**, 733-44 (1976).
28. T. Sellerer, K. Mechlem, R. Tang, K.A. Taphorn, F. Pfeiffer, J. Herzen, Dual-energy x-ray dark-field material decomposition. *IEEE Trans. Med. Imaging* **40**, 974-85 (2021).
29. X. Wu X, H. Liu H, A. Yan, X-ray phase-attenuation duality and phase retrieval. *Opt. Lett.* **30**, 379-81 (2005).
30. S.V. Naydenov, V.D. Ryzhikov, Multi-energy techniques for radiographic monitoring of chemical composition. *Nucl. Instrum. Meth. Phys. Res. A* **505**, 556-8 (2003).
31. A. Astolfo, I.G. Haig, D. Bate, A. Olivo, P. Modregger, Increased material differentiation through multi-contrast x-ray imaging: a preliminary evaluation of potential applications to the detection of threat materials. *arXiv preprint arXiv:2212.11525* (2022).
32. K. Wells, D. Bradley, A review of X-ray explosives detection techniques for checked baggage. *Appl. Radiat. Isot.* **70**, 1729–46 (2012).
33. S.U. Khan, I.U. Khan, I. Ullah, N. Saif, I Ullah, A review of airport dual energy x-ray baggage inspection techniques: image enhancement and noise reduction. *J. X-Ray Sci. Technol.* **28**, 481-505 (2020).
34. A. Shevchuk, J.P.O. Evans, A.J. Dicken, F. Elarnaut, D. Downes, S.X. Godber, K.D. Rogers, Combined x-ray diffraction and absorption tomography using a conical shell beam. *Opt. Exp.* **27**, 21092-101 (2019).
35. A. Alzubaidi, J. Zhang, A.J. Humaidi, A. Al-Dujaili, Y. Duan, O. Al-Shamma, J. Santamaría, M.A. Fadhel, M. Al-Amidie, L. Farhan, Review of deep leaning: concepts, CNN architectures, challenges, applications, future directions. *J. Big Data* **8**, 53 (2021).
36. S. Akcay, T.P. Breckon, Towards automatic threat detection: A survey of advances of deep learning within X-ray security imaging. *Pattern Recognit.* **122**, 108245 (2022).
37. A. Mouton, T.P. Breckon, A review of automated image understanding within 3D baggage computed tomography security screening. *J. X-Ray Sci. Technol.* **23**, 531–555 (2015).
38. S. Shankar, D. Robertson, Y. Ioannou, A. Criminisi, R. Cipolla, Refining architectures of deep convolutional neural networks. In: *Proceedings of the 29th IEEE Conference on Computer Vision and Pattern Recognition*, 7780612, 2212–20 (2016).
39. S.G. Azevedo, H.E. Martz Jr., M.B. Aufderheide III, W.D. Brown, K.M. Champley, J.S. Kallman, G.P. Roberson, D. Schneberk, I.M. Seetho, J.A. Smith, System-Independent Characterization of Materials Using Dual-Energy Computed Tomography. *IEEE Transactions On Nuclear Science*, **63**, 341–349, (2016).

40. D. Shoukroun, L. Massimi, M. Endrizzi, D. Bate, P. Fromme, A. Olivo, Composite porosity characterization using x-ray edge illumination phase contrast and ultrasonic techniques. *Proc. SPIE* **11593**, 115932M (2021).
41. C. Szegedy, V. Vanhoucke, S. Ioffe, J. Shlens, Z. Wojna, Rethinking the inception architecture for computer vision. In: *Proceedings of the IEEE conference on computer vision and pattern recognition* 2818–2826 (2016).
42. A. Krizhevsky, I. Sutskever, G.E. Hinton, Imagenet classification with deep convolutional neural networks. In: *Advances in neural information processing systems* 1097–1105 (2012).
43. A. Astolfo, I. Buchanan, T. Partridge, G.K. Kallon, C.K. Hagen, P.R.T. Munro, M. Endrizzi, D. Bate, A. Olivo, The effect of a variable focal spot size on the contrast channels retrieved in edge-illumination x-ray phase contrast imaging. *Sci. Rep.* **12**, 3354 (2022).
44. S. Ioffe, C. Szegedy, Batch normalization: Accelerating deep network training by reducing internal covariate shift. In: *International conference on machine learning* 448-456 (2015).
45. A.F. Agarap, Deep learning using rectified linear units (relu). *arXiv preprint arXiv:1803.08375* (2018).

**Supplementary Materials**

| No. | Material | No. | Material | No. | Material |
|---|---|---|---|---|---|
| 1 | Granulated Sugar | 26 | Yeast | 51 | Adzuki Beans |
| 2 | Icing Sugar | 27 | Cumin Seeds | 52 | Rubber |
| 3 | Plain Flour | 28 | Sesame Seeds | 53 | Sodium Percarbonate |
| 4 | Wholemeal Flour | 29 | Dried Chickpeas | 54 | Urea Crystals |
| 5 | Ground Almonds | 30 | Raisins | 55 | Blu Tack |
| 6 | Salt | 31 | Oats | 56 | Crushed Chilies |
| 7 | Rock Salt | 32 | Modelling Clay | E1 | PE8 |
| 8 | Cocoa | 33 | Candle Wax | E2 | SX2 Sheet |
| 9 | Cornflour | 34 | Soda Crystals | E3 | RDX/TNT G040 |
| 10 | Coffee Granules | 35 | Simulant Semtex | E4 | Semtex A1 |
| 11 | Ground Coffee | 36 | Epsom Salts | E5 | Blackpowder (Cannon) |
| 12 | Marzipan | 37 | Sponge | E6 | 7.62 Blackpowder |
| 13 | Ready Mixed Icing | 38 | Cotton Wool | E7 | Blackpowder |
| 14 | Quinoa | 39 | Cotton Cloth | E8 | 105 Charge |
| 15 | Rice | 40 | Polyester Cloth | E9 | TNT Flake |
| 16 | Cous Cous | 41 | Notepads* | E10 | Detasheet |
| 17 | Flaxseed | 42 | Playing Cards* | E11 | Ammonal |
| 18 | Lentils | 43 | Post It Notes* | E12 | RDX Class 5 Wet |
| 19 | Bulgar Wheat | 44 | Pencil Eraser – PVC | E13 | Prima Sheet |
| 20 | Tea Leaves | 45 | Tile | E14 | Vihtavuori N560 Powder |
| 21 | Washing Powder | 46 | Soap | E15 | PE4 |
| 22 | Bicarbonate of Soda | 47 | Sponge Cloth | E16 | Blackpowder Type A3 |
| 23 | Dried Oregano | 48 | Iron Tablets | E17 | C4 |
| 24 | Paprika | 49 | Baby Powder (Talc) | E18 | Bullseye Powder |
| 25 | Black Peppercorns | 50 | Petroleum Jelly | E19 | PETN Wet |
|  |  |  |  | 0 | Empty Box |
| C1 | Facewipes | C10 | Handcream | C19 | Socks |
| C2 | Notebook | C11 | Sponge | C20 | Comb |
| C3 | Nappies | C12 | Deodorant | C21 | Lip Balm |
| C4 | Flannel | C13 | Breath Freshener | C22 | Adapter |
| C5 | Tampons | C14 | Hairbrush | C23 | Paracetamol |
| C6 | Cotton Buds | C15 | Hand Sanitiser | C24 | Tape |
| C7 | Toothpaste | C16 | Body Puff | C25 | Floss |
| C8 | Nailbrush | C17 | Gum | C26 | Sewing Kit |
| C9 | Pocket Tissues | C18 | Toothbrush | C27 | Dummy |

**Supplementary Table 1: Materials list.** All materials investigated in the described experiment are listed in the above table, separated into benign materials (1-56), threat materials (E1-19) and cluttering objects (C1-27). The latter were chosen as a combination of items commonly carried in hand luggage, but with the deliberate inclusion of objects creating significant USAXS signal such as facewipes, notebooks, nappies.etc (due to their fibrous microstructure), to make the detection of the underlying material more challenging. Most (threat and benign) materials were placed in plastic boxes, except those marked with an asterisk (*).

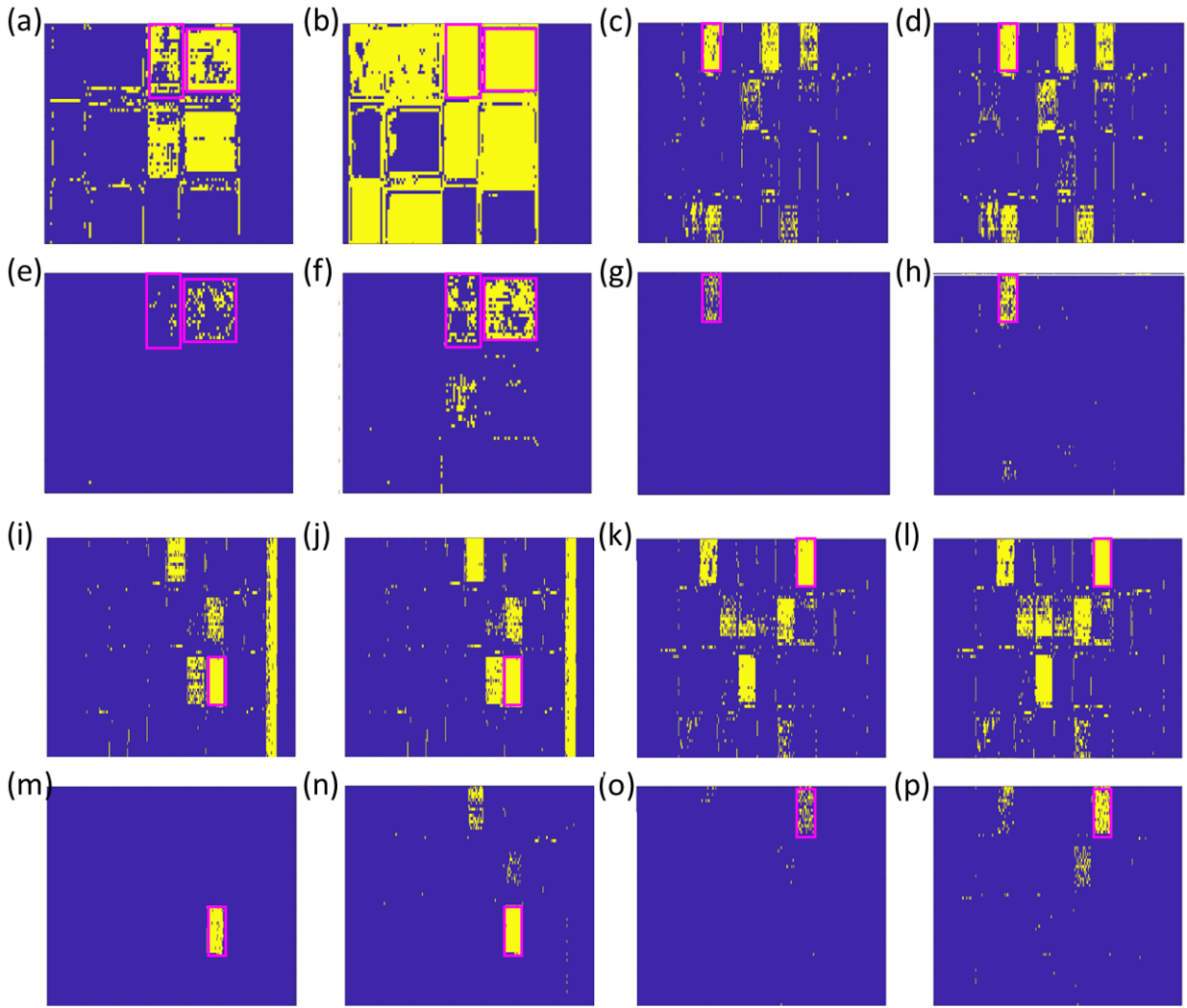

Supplementary Figure 1. Further examples of material discrimination using the different contrast criteria, for TNT Flake (a, b, e, f), Ammonal (c, d, g, h), modelling clay (i, j, m, n) and paprika (k, l, o, p). In each image a pixel is either flagged as part of the material (yellow) or not (blue). Each material of interest is highlighted by a pink box in the image, with TNT Flake having two thicknesses, 12 and 18 mm, present. Images (a, c, i, k) and (b, d, j, l) are based on $Z_e$ alone with the former representing the low threshold cutoff and so a high specificity while the latter high sensitivity. By making use of multiple contrast channels, images (e, g, m, o), required $Z_e$ to be triggered alongside 3 of the other 4 contrast channels for a pixel to be flagged as belonging to a certain material, thus signifying high specificity. Conversely, "high sensitivity" multiple contrast images, (f, h, n, p) simply require that 3 out of 5 channels are triggered, regardless of $Z_e$. As described in the main text, the $Z_e$ only images in either mode (a-d) and (i-l) can only provide so much discrimination and flag other materials as well, but the threshold allows some amount of control over how much is flagged. This is also shown to be true for the multiple contrast images, where the additional information provided by the extra contrast channels allows the focusing of flags to just the material of interest. Here though, the two conditions have some trade-offs, where the high specificity typically leaves just the material alone, but can remove a larger proportion of that too, as seen in (e) and (o), while the high sensitivity leaves significantly more of each material pixels flagged but with a smaller proportion of other materials being flagged too, which could potentially lead to false positives if used in a security application. It should be noted that other conditions could be chosen, particularly in the multiple contrast, as well as tighter or looser bounds on each channel. As seen in Fig. 1 of the main article, the different contrast channels offer varying separation.

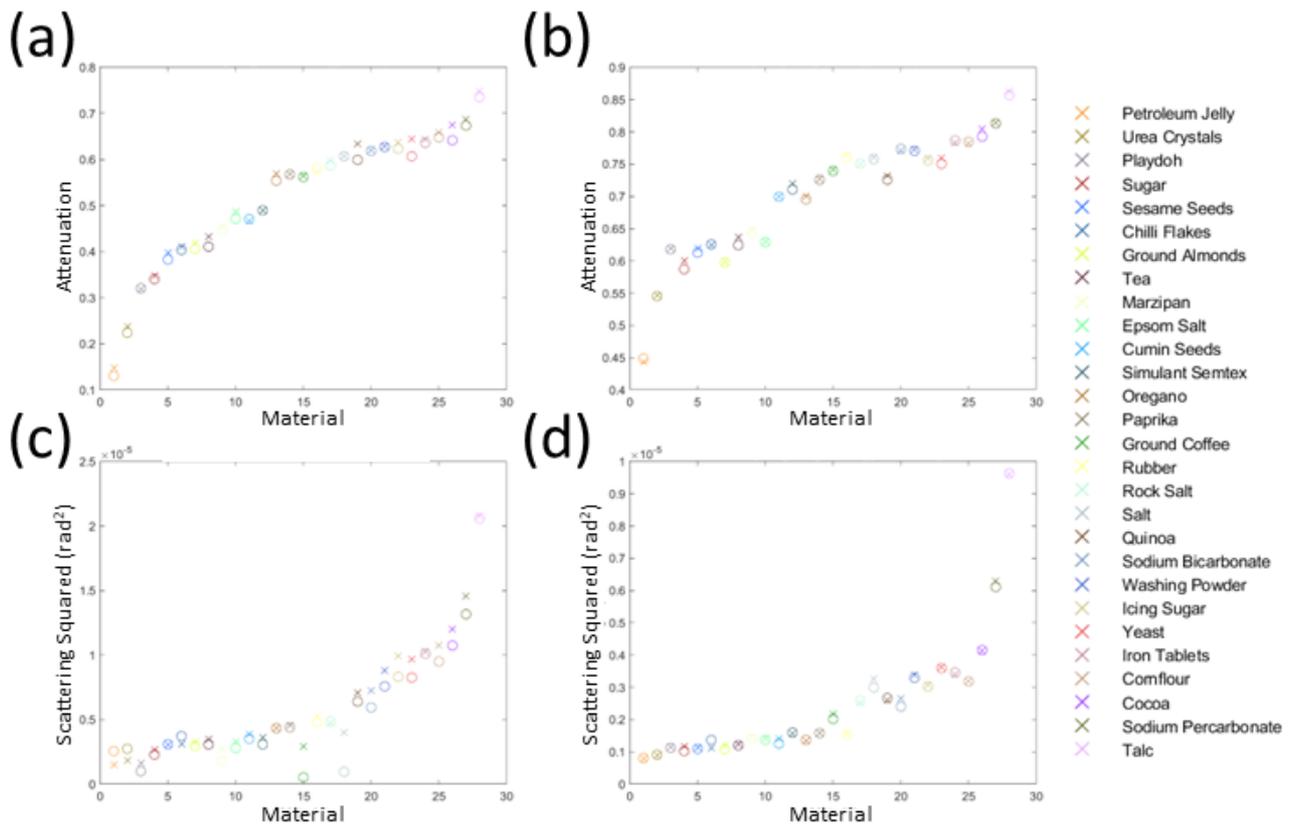

Supplementary Figure 2. Overlap of cluttering items can be handled by correcting the values using an area of the cluttering item clear of overlap. In this case, the items are obscured by a laptop bag, which primarily attenuates and scatters. The crosses on the figures represent the material only values, while the circles show values of the obscured materials post correction. (a) and (b) are the low and higher energy attenuation, while (c) and (d) are the low and high energy dark field signals. From this it is clear that the correction works well for the majority of materials, with the only real exception being the salt and rock salt in (c), the low energy dark field. This is thought to be because they are materials with large grains (i.e., significant refraction), so may be better corrected with the standard deviation of phase; again this relates to potentially giving different weights to different contrast channels, which will be explored in future work.

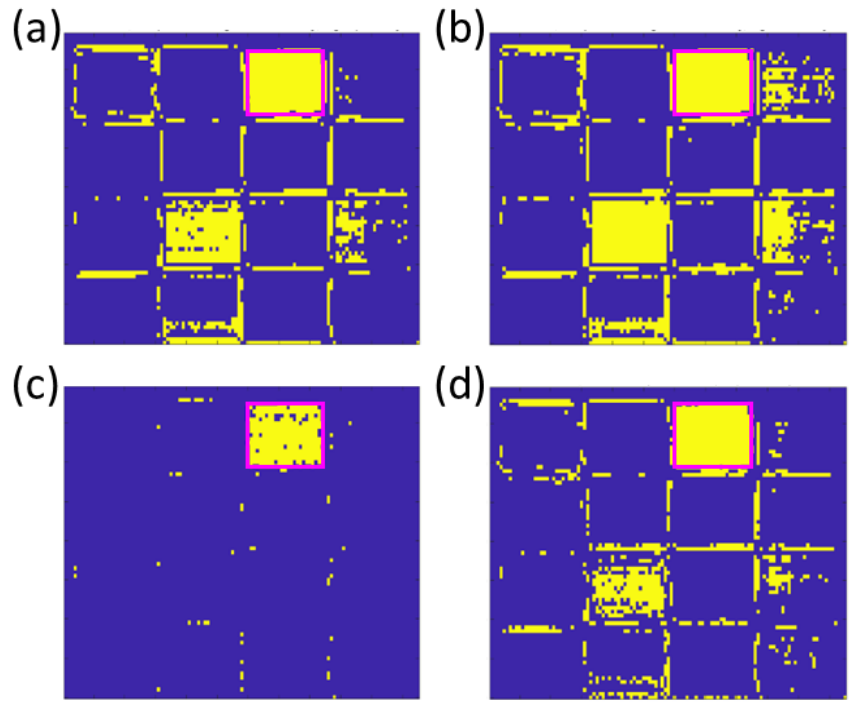

Supplementary Figure 3. Example of the material discrimination for washing powder with overlapping clutter. In this case, all materials have been obscured by a notebook and the correction applied prior to selection using the process described in the previous supplementary figure and in the main text. As before, (a) and (b) are based on $Z_e$ alone (a) representing the low threshold cutoff and so a high specificity, while (b) represents the high sensitivity mode. Images (c) and (d) once again make use of multiple contrast channels with (c) requiring $Z_e$ to be triggered alongside 3 of the other 4 contrast channels, thus signifying "high specificity", while (d) simply requires that 3 out of 5 channels are triggered, regardless of $Z_e$, giving the "high sensitivity" mode. Once again, the $Z_e$ only images, (a) and (b), highlight all pixels of the washing powder but even at "high specificity" also flag a number of other materials, especially cocoa. Using multiple channels, (c) and (d), cleans this up, particularly in the high specificity case where nearly all the washing powder is flagged but none of the cocoa, showing the effectiveness of the correction.

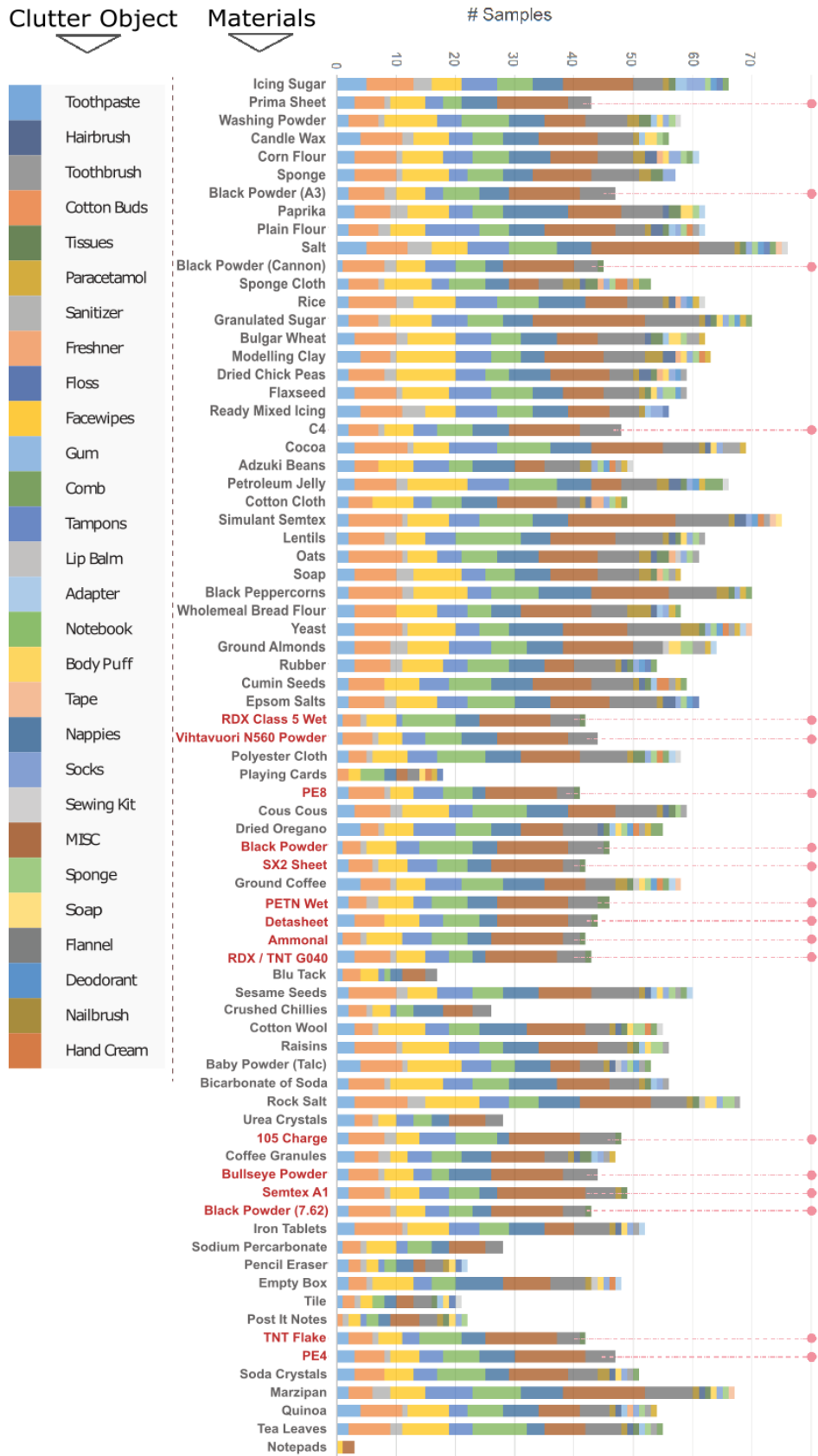

**Supplementary Figure 4. Characteristics of the Deep Learning Dataset,** based on the *Materials* and *Cluttering Objects* listed in *Supplementary Table I*. In the column on the far left, the figure lists the cluttering object classes; each class is coded through a certain color.

Note that our most successful deep learning experiment comes from a hierarchical approach (see Supplementary Material 6), where in the first stage we classify input images for the (cluttering) object classes they contain, and in the second stage a separate network is trained pertaining to each object class, to classify threat and non-threat materials (obscured by the respective cluttering object class). While this may seem a non-scalable method (since the object classes obscuring the materials need to be known a-priori), we emphasize that, out of 27 cluttering objects, we trained/tested with only 7 classes, by specifying only 6 objects (cotton buds, face wipes, tampons, notebooks, nappies, flannel) and grouping everything else into a "miscellaneous" class. The miscellaneous class therefore learns mixed object features, and any new data coming without specific object information may be simply seen as miscellaneous. Basically, our experiment showed that, so long as training is done on a subset of specific objects, most others can be trained under a single mixed class, which significantly helps the method's scalability.

The second column lists all threat/non-threat materials, and the color bars to the right of each material show in how many instances each of these is obscured by the different object classes, using the color-coding shown in the first column. For example, the dataset features 67 images that contain icing sugar (see top row). Out of these, in five cases icing sugar is obscured by toothpaste, in eight by cotton buds, and so on. In cases where icing sugar is obscured by both toothpaste and cotton buds, two entries are made in the above figure, and we also duplicate such images (with a mild data-augmentation style modification) for training. Note that in the figure we have singled out some of the objects that were included in the miscellaneous class, and indicated the remaining ones as MISC. This is done to illustrate that these other objects are weakly represented individually, but they may make a substantial mixed class when taken together; we iterate that, this notwithstanding, the training/testing is done only with the samples of the seven classes mentioned above (six specific objects plus miscellaneous). The red pins on the far right of the figure highlight all the explosive materials, all of which appear in at least 40 images, typically more.

Though not shown in the figure, it is worth mentioning that we split the dataset to maintain nearly the same train and test set distribution. This implies that the relative proportion of the object classes obscuring a given threat/non-threat material remains similar, both while training and testing.

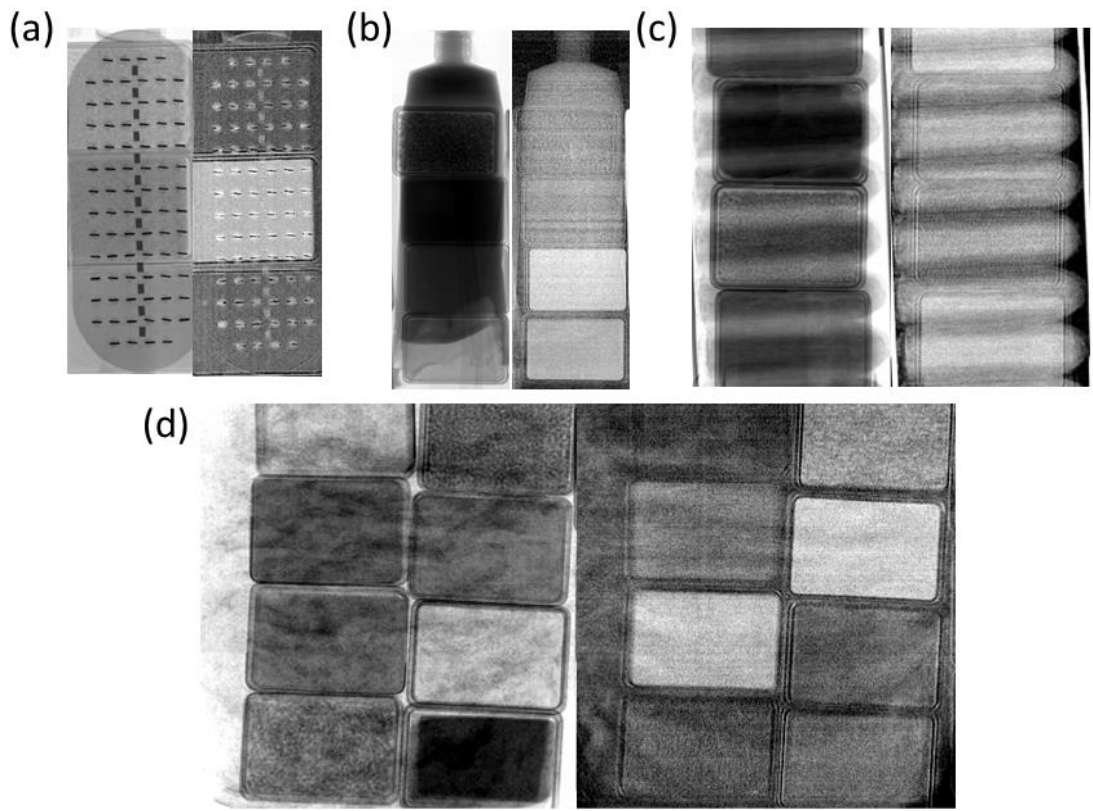

**Supplementary figure 5**: Example images from the trial, in each case showing an attenuation (left) and dark-field image (right) side by side: their complementarity is apparent, with the materials often standing out much more clearly against the cluttering objects in the dark field images, and contrast often switching polarity from one modality to the other (materials that look dark in attenuation turning bright in dark field and vice-versa). Panels (a-d) show boxes containing various materials being obscured by a nailbrush, toothpaste, tampons and nappies, respectively. It is worth noting that many of the clutter items add randomness to the images, particularly over the course of many days of scanning. Examples include the location of bristle overlap for the nailbrush, amount of toothpaste in different parts of the tube, and variation in overlap in the case of discrete and non-homogeneous items like the tampons.

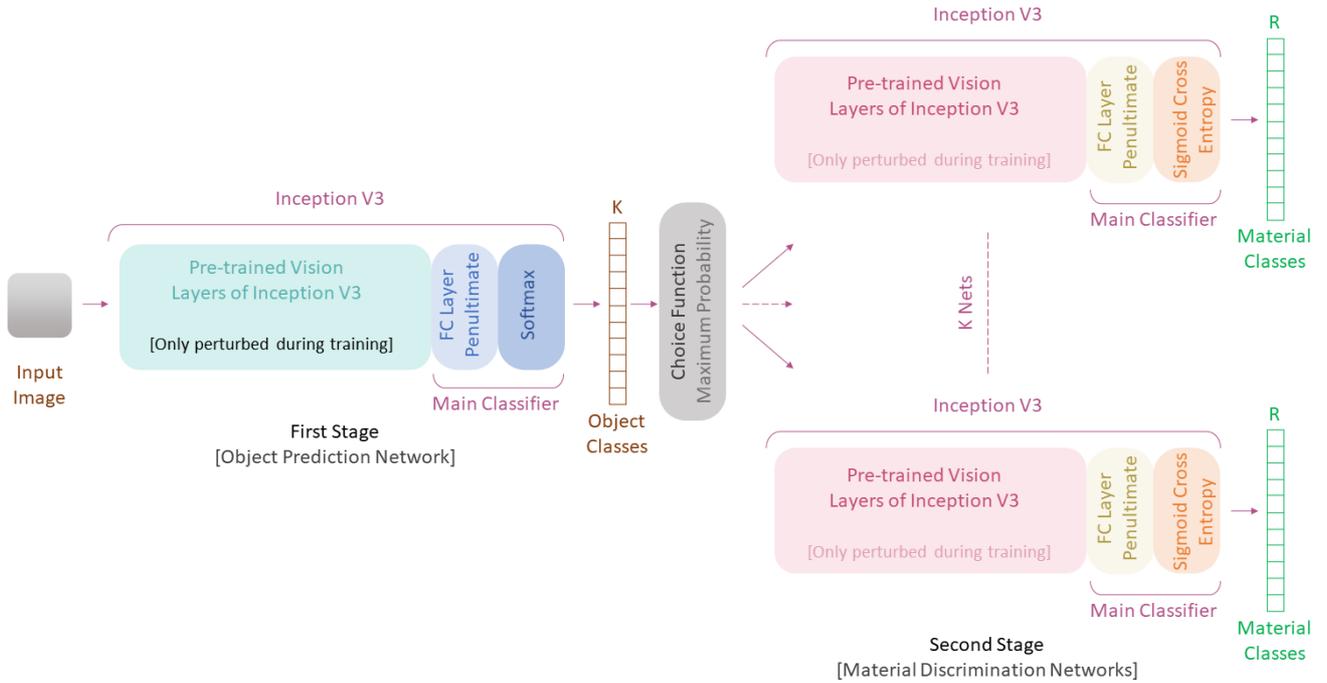

**Supplementary Figure 6**: **Hierarchical Deep Net Architecture.** The figure shows the hierarchical (two-stage) architecture, which provided the best results. The first stage consists of a single network segregating $K = 7$ cluttering object classes, while the second stage features $K$ individual networks, each separating $R = 75$ threat/non-threat materials, obscured by the corresponding object classes.

Inception-V3 (41) has been used as the base architecture, pre-trained on ImageNet (42). As part of the vision layers, Inception-V3 contains multiple Convolutional and Max Pooling layers at the start, followed by topologically-varying Inception Modules, which are interlaced with Grid Size Reduction mechanisms. Batch Normalization (44) and ReLU (45) are applied after all the convolutional layers. The penultimate Fully Connected (FC) layer in the Main Classifier is preceded by an Average Pool and a Dropout layer. Inception-V3 uses an Auxiliary Classifier as a regularizer to avoid the problem of vanishing gradients, whose training loss is added to the training loss of the main classifier with a weight factor of 0.4. The auxiliary classifier also contains an FC layer, penultimate to its loss function layer, and preceded by two convolutional layers. Since the auxiliary classifier is used only during the training phase, and prediction is done through the main classifier only, the former is not shown in the above figure for the sake of clarity.

Due to the limited amount of available training data, we used transfer learning on pre-trained Inception-V3 models. Specifically, we re-trained the penultimate FC layers of both the main and the auxiliary classifiers with a learning rate equal to 10 times that of the pre-trained vision layers and the pre-trained convolutional layers in the auxiliary classifier. This allows for the vision layers to get slightly perturbed in their parameters (if need be), which imparts better generalization capability to the network, in comparison to the case where vision layers may be completely frozen (i.e., not allowed to participate in training at all). Also, the training of our two-stage network is done end-to-end, but only after the object detection network (of the first stage) is trained to a reasonable accuracy; this provides better convergence.

The first stage of the architecture uses a softmax function for classification to detect only the "prevailing" cluttering object i.e., the one that most obscures a given material. Since materials may be obscured by more than one object, we could have used a sigmoid cross-entropy loss instead, set a probability threshold of choosing multiple object classes, and then combined the outputs from multiple corresponding networks in the second stage; however, this seemed excessive for such a small number of cluttering object classes. All second-stage networks use the sigmoid cross-entropy loss. During inference, the sigmoid function is applied over the outputs of the penultimate layers. Here, a multi-label loss is important since a threat material is almost always present in an image alongside several non-threat materials, and all are expected to be detected. Also, an image may contain two or more threat materials, and once the obscuring object class has been detected, all these threat materials should ideally be identified (in the second stage).

**Authors contributions**
AO and DB conceived research. TP, SS, IB, PM, AA, AO performed research. SS, TP, IB analyzed data; AO and TP wrote the paper.

**Competing interests**
DB is a Nikon employee; AA was a Nikon employee at the time this research was conducted. AO is a named inventor on patents held by UCL protecting the x-ray imaging technology used to obtain the results presented in this paper.

**Data sharing plans**
The datasets generated and/or analysed during the current study are not publicly available on grounds of security but are available from the corresponding author on request. Computer code used to perform phase retrieval is available from AA. Machine learning code is available from SSS. Interested readers are invited to contact the corresponding author.

**Funding information**
This work was supported by the EPSRC (grant EP/T005408/1) and by the Innovative Research Call in Explosives and Weapons Detection. The latter is a Cross-Government programme sponsored by a number of Departments and Agencies under the UK Government's CONTEST strategy in partnership with the US Department of Homeland Security, Science and Technology Directorate (grant 84887-537251). A.O. was supported by the Royal Academy of Engineering under the Chairs in Emerging Technologies scheme (grant CiET1819/2/78).